\documentclass[useAMS,usenatbib]{mn2e}
\usepackage{graphicx,amsmath,multirow,amssymb}
\usepackage{natbib}
\newcommand{\comment}[1]{}

\def\simgt{\lower.5ex\hbox{$\; \buildrel > \over \sim \;$}}
\def\simlt{\lower.5ex\hbox{$\; \buildrel < \over \sim \;$}}

\title{AGB and SAGB stars: modelling dust production at solar metallicity}
\author[Dell'Agli et al.]{F. Dell'Agli$^{1,2}$, D. A. Garc\'{\i}a-Hern\'andez$^{1,2}$, R. Schneider$^3$, P. Ventura$^3$,  
\newauthor
 F. La Franca$^{4}$, R. Valiante$^{3}$, E. Marini$^{4}$, M. Di Criscienzo$^{3}$\\
$^{1}$Instituto de Astrof\'{\i}sica de Canarias, V\'{\i}a L\'actea s/n, E-38205 La Laguna, Tenerife, Spain \\
$^{2}$Departamento de Astrof\'{\i}sica, Universidad de La Laguna (ULL), E-38206 La Laguna, Spain\\
$^3$INAF -- Osservatorio Astronomico di Roma, Via Frascati 33, 00040, Monte Porzio Catone (RM), Italy \\
$^4$Dipartimento di Matematica e Fisica, Universita degli Studi `Roma Tre', Via della Vasca Navale 84, I-00146 Roma, Italy}

\begin{document}

\date{Accepted, Received; in original form }

\pagerange{\pageref{firstpage}--\pageref{lastpage}} \pubyear{2012}

\maketitle

\label{firstpage}

\begin{abstract}

We present dust yields for asymptotic giant branch (AGB) and super--asymptotic giant branch 
(SAGB) stars of solar metallicity. Stars with initial mass $1.5~M_{\odot} \leq M_{\rm ini} \leq 3~M_{\odot}$ reach the carbon star stage during the AGB phase and produce mainly solid carbon and SiC. 
The size and the amount of the carbon particles formed follows a positive trend with the
mass of the star; the carbon grains with the largest size ($a_{\rm C} \sim 0.2\mu$m) 
are produced by AGB stars with $M_{\rm ini} = 2.5-3~M_{\odot}$, as these stars are those
achieving the largest enrichment of carbon in the surface regions. The size of
SiC grains, being sensitive to the surface silicon abundance, 
keeps around $a_{\rm SiC} \sim 0.1\mu$m. The mass of carbonaceous dust formed 
is in the range $10^{-4} - 5\times 10^{-3}~M_{\odot}$, whereas the amount of SiC produced
is $2\times 10^{-4} - 10^{-3}~M_{\odot}$.
Massive AGB/SAGB stars with $M_{\rm ini} > 3~M_{\odot}$ experience HBB, that inhibits 
formation of carbon stars. The most relevant dust species formed in these stars are silicates and
alumina dust, with grain sizes in the range $0.1\mu m < a_{\rm ol} < 0.15\mu$m 
and $a_{\rm Al_2O_3} \sim 0.07\mu$m, respectively. The mass of silicates produced spans the
interval $3.4\times 10^{-3}~M_{\odot} \leq M_{\rm dust} \leq 1.1\times 10^{-2}~M_{\odot}$
and increases with the initial mass of the star. 

\end{abstract}


\section{Introduction}

The recent years have witnessed a growing interest towards the post core
helium-burning evolution of low- and intermediate-mass stars ($M < 8~M_{\odot}$). These 
late evolutionary phases, the AGB or SAGB (for initial masses $6~M_{\odot} < M < 8~M_{\odot}$, 
which develop a O-Ne core), despite representing only a few percent of the overall stellar
life, are of fundamental importance to understand the feedback of these objects
on the host environment. This is because AGB and SAGB stars are characterised by
strong mass-loss rates, which favour the loss of the whole external envelope,
before the beginning of the planetary nebula and white dwarf evolution.

During the AGB phase the stars eject gas with a chemical composition altered by
internal nucleosynthesis processes, thus enriching the
interstellar medium. Several theoretical studies have presented
mass- and metallicity-dependent AGB stellar yields 
(see, among others, \citealt{cristallo09, cristallo15, doherty14, karakas10,
karakas14a,  karakas14b, karakas16, ventura13, ventura14, dicrisci16}). These
works outlined several similarities but also significant differences in the
predicted yields. These are mostly due to the different description of convection, both in
terms of the efficiency of convective transport and of the treatment of the
convective/radiative interface.

A strong interest towards AGB stars is also motivated by the thermodynamical
structure of the circumstellar envelope, which proves to be an environment extremely
favourable to the condensation of gaseous molecules into dust particles. The
surface layers of these stars are sufficiently cool (${\rm T_{eff}} < 4000$K)
to allow dust formation at typical  distances of 3--10 stellar radii from the
surface, where the densities are large enough 
\citep[$\rho > 10^{-14} \rm{gr/cm^3}$,][]{gs85} to allow the formation of 
meaningful quantities of dust. 

The pioneering investigations by the Heidelberg group \citep{fg01, fg02, fg06}
set up the framework to model dust formation in the circumstellar envelope of
AGB stars: the schematisation adopted is based on the assumption that the wind
expands isotropically from the stellar surface, under the push of radiation
pressure, acting on the newly formed dust grains. While interesting improvements
to this basic treatment, accounting for the effects of shocks coupled to the AGB
evolution are in progress \citep{marigo16}, this approach is at present the
only way to allow the description of dust formation in the winds of
stars of different mass and chemical composition, extended over the whole AGB
phase. Indeed, different research groups have modelled the formation and growth
of solid particles in the wind of low-metallicity AGB stars, calculating the
dust yields from this class of objects \citep{paperI, paperII, ventura14,
dicrisci13, nanni13, nanni14}. These models have been successfully used to
interpret the near- and mid-infrared observations (mainly obtained with the {\it
Spitzer Space Telescope}) of evolved stars in the Magellanic Clouds
\citep{flavia14b, flavia15a, flavia15b, nanni16} and in the Local Group (LG)
galaxy IC 1613 \citep{flavia16}. An important outcome of these investigations
was the characterization of the individual sources observed, in terms of mass,
formation epoch and chemical composition of their progenitors. Furthermore,
these studies allowed to infer important information on the past history of the
host galaxy (particularly the star formation history and the mass-metallicity
relationship) and provided valuable constraints on the physical mechanisms
relevant to understand the main evolutionary properties of AGB stars.

As we have mentioned above, these works have been so far based on
low-metallicity extra-galactic AGB stars. This is partly due to the fact that
the distances to Galactic sources are basically unknown or very uncertain,
so that interpreting the observations is quite difficult. This limitation
will be largely overcome by the results of the Gaia
mission, because the parallaxes - so the distances and luminosities - will be measured with high
accuracy for an important fraction of Galactic AGB stars, including those
exhibiting a large degree of obscuration. This will open new frontiers in the
AGB field, because the sample of Galactic AGB stars is much larger and
extends to a wider metallicity range than the AGB population of the Magellanic
Clouds and other LG galaxies.

In order to be ready for the Gaia challenge, we have recently extended the AGB
models, so far limited to metallicities $Z \leq 8\times 10^{-3}$, to solar
metallicity (Di Criscienzo et al. 2016, hereinafter DC16). In DC16 we presented
the main evolutionary properties of the solar metallicity AGB models as well as 
the chemistry of the gas ejected in the interstellar medium. Here we focus on
the dust formation properties and present new dust yields based on a
self-consistent coupling of the wind description with the DC16 AGB evolutionary
models. In addition, these solar-metallicity AGB dust models will be essential
to interpret the rich dataset of IR photometric observations (e.g., from space
missions like MSX, AKARI and WISE) of Galactic sources as well as the already
available {\it Spitzer} photometric observations of evolved stars in nearby
high-metallicity galaxies such as M 31 \citep{mould08}, M 33 \citep{mcquinn07, javadi11a, javadi11b, javadi15} and M 32 \citep[e.g.][]{jones15}. 
The results presented here and in our previous papers of this
series  \citep{paperI,
paperII, dicrisci13, ventura14}, complemented by a new grid of mass- and metallicity-dependent SN
dust grid (\citealt{marassi2014, marassi2015,bocchio2016}, Marassi et al. in prep) will allow to fully
assess the role of AGB stars as cosmic dust polluters, in present-day galaxies \citep{schneider2014}
as well as at very high redshifts \citep{valiante2009, valiante11, mancini2015, mancini2016}.


\section{Dust formation model}
\label{dustmod}

We describe the growth of dust particles in the circumstellar envelope of AGB
stars by means of the schematisation introduced by the Heidelberg group 
\citep{fg01, fg02,  fg06}. We provide here a short description of the basic concepts upon which the modelling of the wind dynamics and of the dust formation process is based; we address the 
interested reader to the previous works on this argument \citep{paperI,
paperII, dicrisci13, ventura14, flavia14a}, which include the full set
of equations used and a more exhaustive discussion on this argument.

The wind is assumed to expand isotropically from the
surface of the star. Dust formation occurs once the growth rate of solid grains
exceeds the rate of destruction (hereafter the decomposition rate). 
The first is determined by the
deposition efficiency of gaseous molecules on the already formed solid
particles. For each dust species, the decomposition rate is found via the 
evaluation of the vapour pressures of individual molecular species involved, 
under  thermodynamic equilibrium conditions \citep{gs99}.

The description of the wind is self-consistently coupled to the results of
stellar  evolution: the input parameters entering dust formation, namely the
mass ($M$),  luminosity ($L$), effective temperature ($T_{\rm eff}$), radius ($R_{\ast}$),
mass loss rate  ($\dot M$) and surface chemical composition of the star, are the
natural output of AGB evolutionary modelling. 

The kind of particles formed in the circumstellar envelope depends on the
surface chemical  composition of the star, mainly on the surface carbon-to-oxygen
(C/O) ratio.
This is because the CO molecule  is extremely stable \citep{sharp90}, thus the
least abundant element between carbon and oxygen is locked into CO molecules. 
Based on stability arguments, we assume that in  carbon--rich environments the
species of dust formed are solid carbon, solid iron and  silicon carbide,
whereas in M stars the formation of silicates (olivine, pyroxene and quartz),
solid iron and alumina dust is accounted for \citep{sharp90, fg06}.
For each dust species it is possible to identify the key element, which is the least abundant among
the various chemical species required to form the solid particles: such key--element is extremely important, as
it determines the largest amount of a specific kind of dust particles which is possible to form. A list of the dust species considered here, with the reactions of formation, the
key--elements and the sticking coefficients adopted, is shown in Table 1 of \citet{ventura14}\footnote{After the analysis published in \citet{flavia14a} where different value of the sticking coefficient for the alumina dust has been explored, we adopted the value 0.1 which is consistent with the recent laboratory measurements by \citet{Takigawa15}.}.

\section{Solar metallicity AGB stars}
\label{agbphys}

The AGB evolutionary models calculated with the code ATON \citep{cm91} and used in the present work have been recently published by DC16. The models span the range of initial masses  $1~M_{\odot}
\leq M_{\rm ini} \leq 8~M_{\odot}$. The upper limit is determined by the core
collapse via electron capture experienced by $M_{\rm ini} > 8~M_{\odot}$ stars.
During the AGB phase the stars evolve on a degenerate core composed of carbon
and oxygen with the exception of the stars with mass above $\sim 6.5~M_{\odot}$,
which form a core composed of oxygen and neon (the so-called SAGB stars),
owing to an off-center ignition of carbon in conditions of partial degeneracy
\citep{garcia1, garcia2, siess06, siess07, siess09, siess10}. The interested
reader is referred to DC16 for an exhaustive discussion on the evolutionary
properties of these stars. Here we only provide a brief summary of the physical
and chemical evolutionary aspects most relevant to the dust formation process
in their winds.

The stars of initial mass $M_{\rm ini} \geq 3.5~M_{\odot}$, to which we will refer
as ``massive AGB/SAGB stars", experience hot bottom
burning (hereinafter HBB) at the base of the outer convective envelope
\citep{renzini81}. The activation of proton-capture nucleosynthesis at the
bottom of the external mantle leads to a considerable increase in the luminosity
of the star \citep{blocker91}, with significant deviations from the classic core
mass - luminosity relation \citep{paczynski}. The largest luminosities reached
range from $L \sim 2\times 10^4 L_{\odot}$, for $M_{\rm ini} = 3.5~M_{\odot}$, to $L
\sim 10^5 L_{\odot}$,  for $M_{\rm ini} = 8~M_{\odot}$. The temperatures at which
the bottom of the surface convective zone is exposed, which affect the degree of
nucleosynthesis experienced, span the range $[30-100]$~MK (see Fig.~1 in DC16). 

\begin{figure*}
\begin{minipage}{0.45\textwidth}
\resizebox{1.\hsize}{!}{\includegraphics{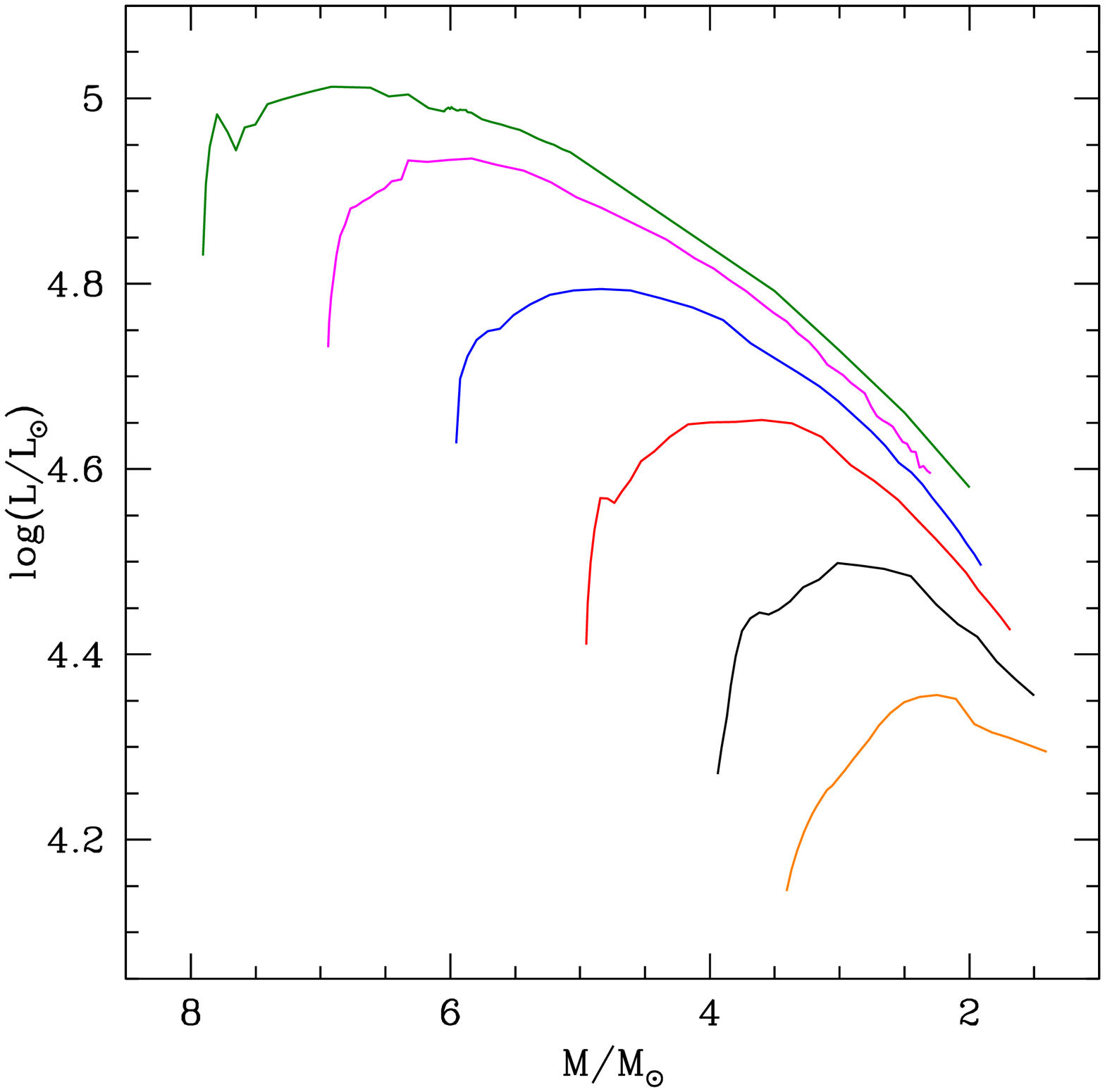}}
\end{minipage}
\begin{minipage}{0.45\textwidth}
\resizebox{1.\hsize}{!}{\includegraphics{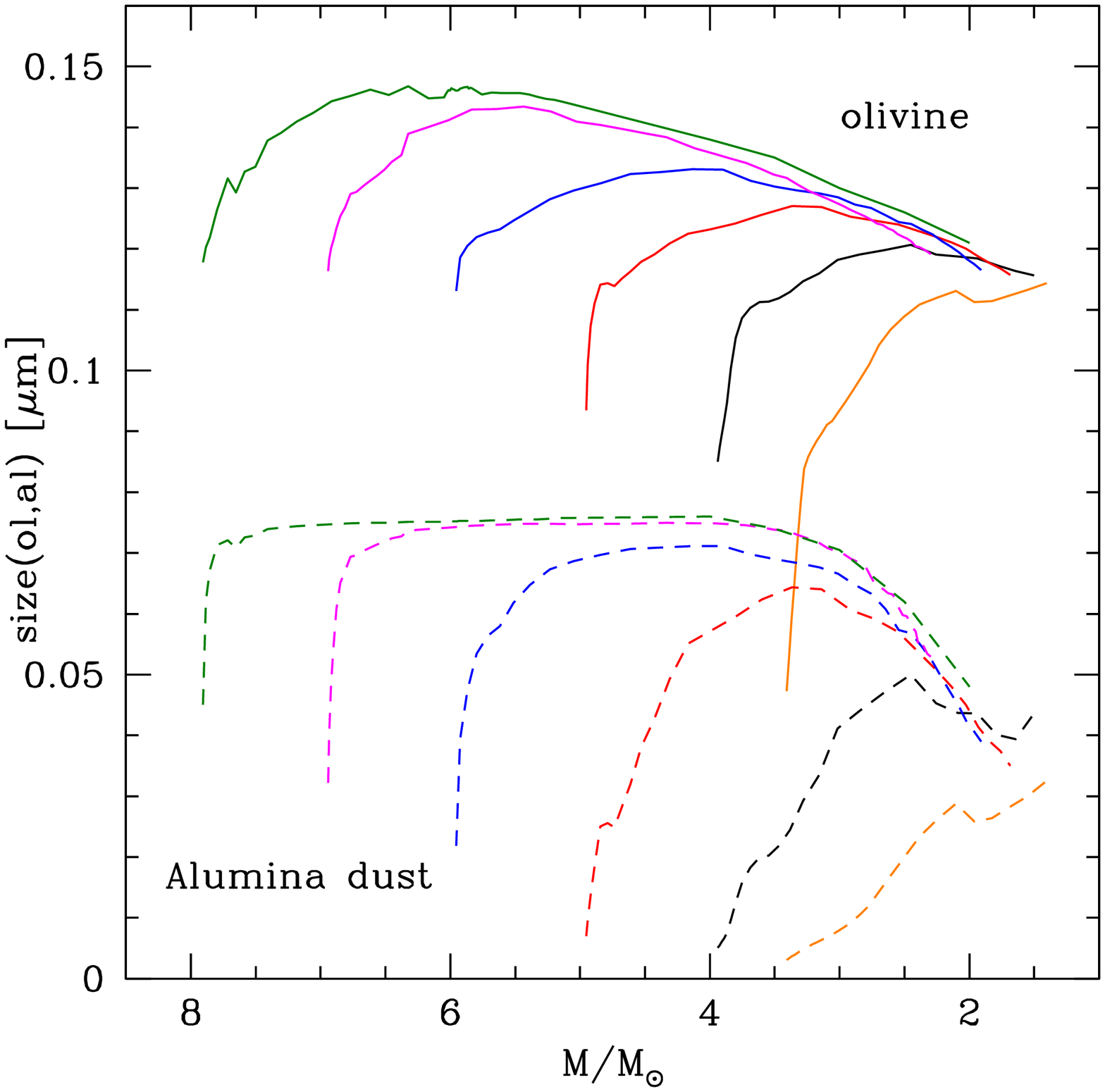}}
\end{minipage}
\vskip-0pt
\begin{minipage}{0.45\textwidth}
\resizebox{1.\hsize}{!}{\includegraphics{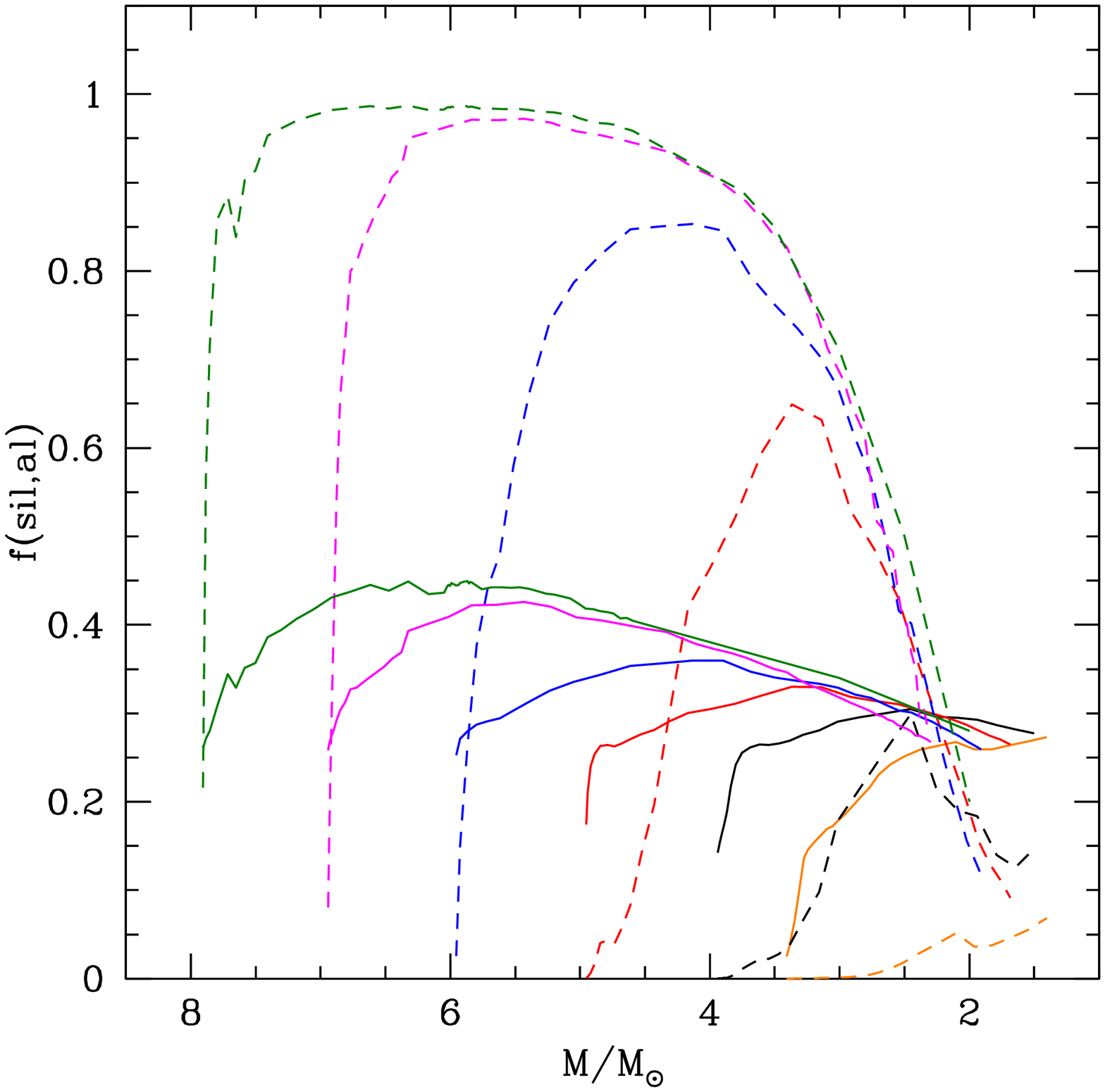}}
\end{minipage}
\begin{minipage}{0.45\textwidth}
\resizebox{1.\hsize}{!}{\includegraphics{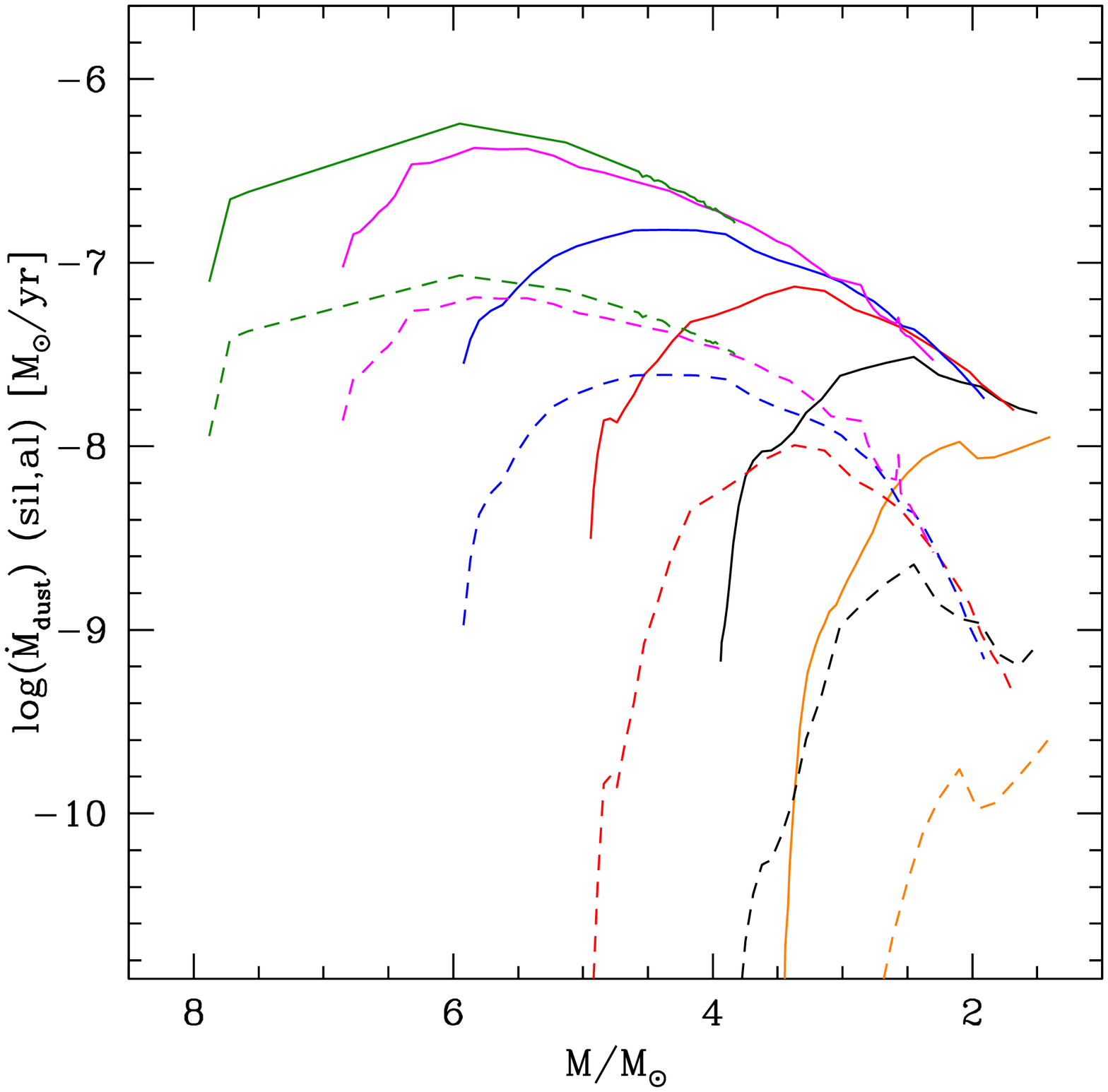}}
\end{minipage}
\vskip+10pt
\caption{The variation during the AGB phase of the physical and dust properties
of massive AGB/SAGB stars, with initial mass in the range $[3.5 - 8]~M_{\odot}$, shown 
by the coloured lines from orange to green, respectively. The current mass of
the star (decreasing during the evolution) is displayed on the abscissa. The
quantities shown in the four panels are: luminosity (top-left), the size of olivine and 
Al$_2$O$_3$ grains (top-right, solid and dashed lines, respectively), the fraction of 
silicon condensed into silicates and the fraction of aluminium condensed into alumina
dust (bottom-left, solid and dashed lines, respectively), the silicates and alumina dust
mass-loss rates (bottom-right, solid and dashed lines, respectively). A coloured version 
of this plot is available online.}
\label{fsil}
\end{figure*}

On the chemical side, massive AGB/SAGB stars never reach the C-star stage,
because the surface carbon is destroyed at the base of the convective envelope
by proton capture reactions, which prevent the achievement of the C/O$ > 1$
condition. Due to the ignition of CN cycling, the ejecta from these stars are
nitrogen-rich (with an overall N increase slightly below a factor of 10 with respect to the initial abundance) and
carbon-poor, with the carbon content being one order of magnitude smaller than
the original gas out of which the stars formed. Unlike lower metallicity AGB
stars, in this case the HBB temperatures are below $100$~MK, which reflects into
a modest (below $\sim 20\%$) depletion of oxygen and magnesium and a negligible
production of aluminium. Sodium is produced in significant quantities, with an
average sodium increase in the ejecta by a factor of $\sim 4$ (see Figs.~8
and 9 in DC16).

The evolution of AGB models with $M_{\rm ini} < 3.5~M_{\odot}$, to which we will
refer to as ``low-mass AGB stars", is very different from that of their higher mass
counterparts. This is because the core mass is $ < 0.8~M_{\odot}$, too small to
 activate HBB (Ventura et al. 2013). In this case, the only physical mechanism
able to alter the surface chemical composition is the third dredge-up (TDU),
which produces a gradual increase in the surface carbon and, eventually, the
formation of a carbon star (C/O$>$1). 

On the physical side, the achievement of the C-star stage causes a considerable
increase in the surface molecular opacities that favours a general cooling and
a considerable expansion of the external regions \citep{marigo02, vm09, vm10}.
The stellar effective temperature ($T_{\rm eff}$) decreases down to $\sim$1900 K, in
conjunction with the maximum surface carbon abundance. Consequently, the
outermost regions of the star become less and less gravitationally bound and
the mass-loss rate increases. After the
C-star stage is reached, the evolutionary times become short and the relative duration of the C-star phase is below
$15\%$ respect to the entire AGB phase (see Table 1 in DC16).\\

The surface chemical composition of low-mass AGB stars is entirely dominated by
the effects of TDU. The gas lost by these stars is enriched in carbon, with
a maximum increase by a factor of $\sim 3$ achieved in the
$M_{\rm ini}=3~M_{\odot}$ model. The final surface carbon abundances range from
$X_{\rm C}=7.3\times10^{-3}$ for $M_{\rm ini}=1.5 \, M_{\odot}$, to $X_{\rm C}=9.7\times10^{-3}$
for $M_{\rm ini} = 3 \, M_{\odot}$. The enrichment in carbon increases with $M_{\rm ini}$
because more massive stars are exposed to a higher number of TDU events before
the envelope is completely lost.

\section{Dust production}

As discussed in Section \ref{dustmod}, the kind of dust grains (i.e., their
specific chemical composition) formed in the wind of AGB stars is mainly
determined by the surface C/O ratio: oxygen-rich stars produce silicates and
alumina (Al$_{2}$O$_{3}$) dust grains, carbon-stars produce solid carbon and silicon carbide
(SiC) grains.

The surface chemical composition of solar metallicity, AGB stars is
extremely sensitive to the initial mass (see Section \ref{agbphys}): stars with
$1.5~M_{\odot} \leq M_{\rm ini}\leq 3~M_{\odot}$ become carbon stars\footnote{As
discussed in DC16, stars of mass below $1.5~M_{\odot}$ never reach the C-star
stage, because they loose the whole external mantle before the surface carbon exceeds
oxygen.}, while their higher mass counterparts evolve as M-stars for the whole
AGB phase. This dichotomy reflects into the dust composition, which is dominated
by silicates in massive AGB/SAGB stars and by solid carbon in the
low-mass domain. In the following, we analyse these two groups separately. We do
not discuss stars with $M_{ini} < 1.5~M_{\odot}$, because
they produce only a negligible amount of silicates during their life.

\begin{figure*}
\begin{minipage}{0.48\textwidth}
\resizebox{1.\hsize}{!}{\includegraphics{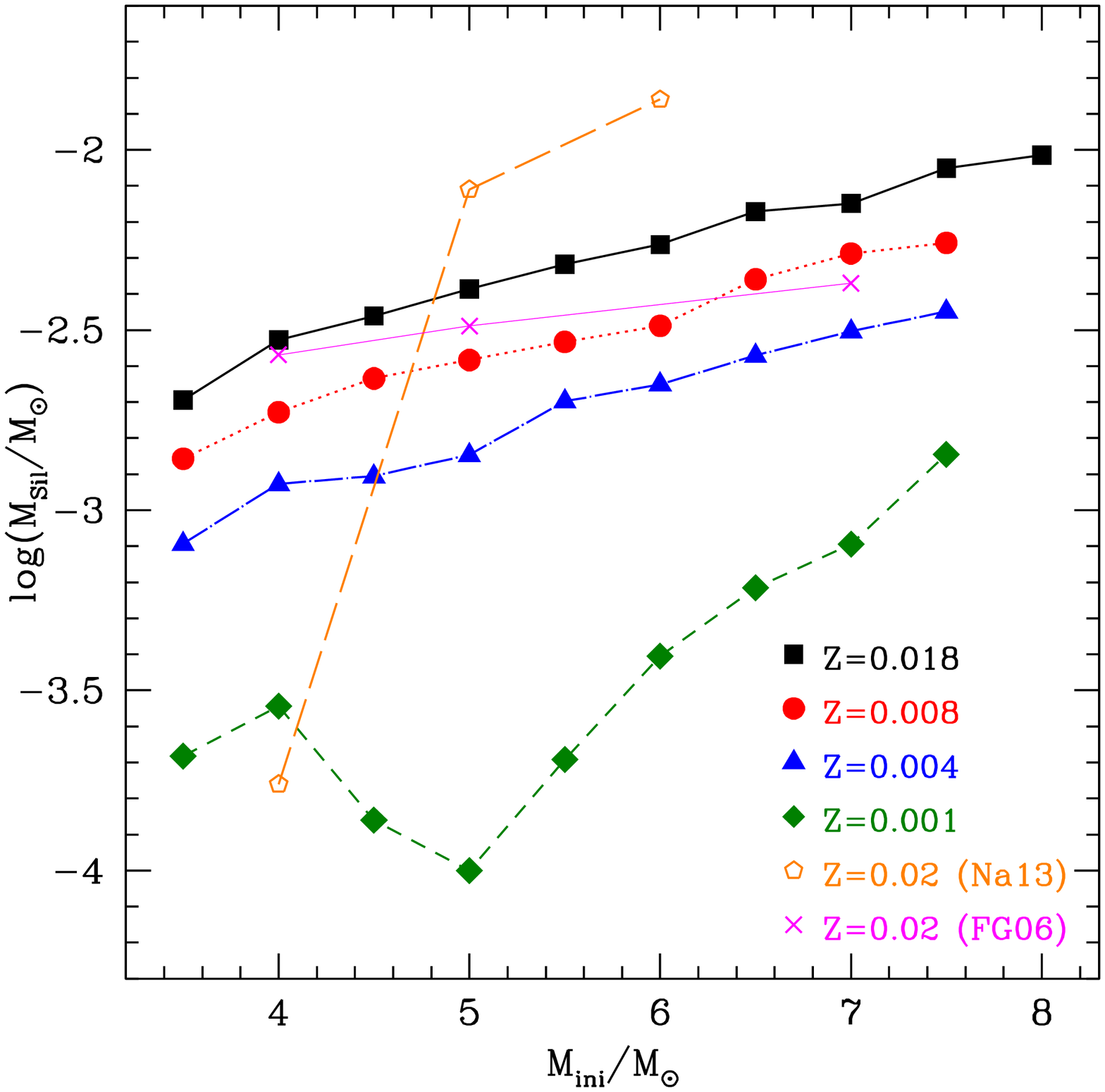}}
\end{minipage}
\begin{minipage}{0.48\textwidth}
\resizebox{1.\hsize}{!}{\includegraphics{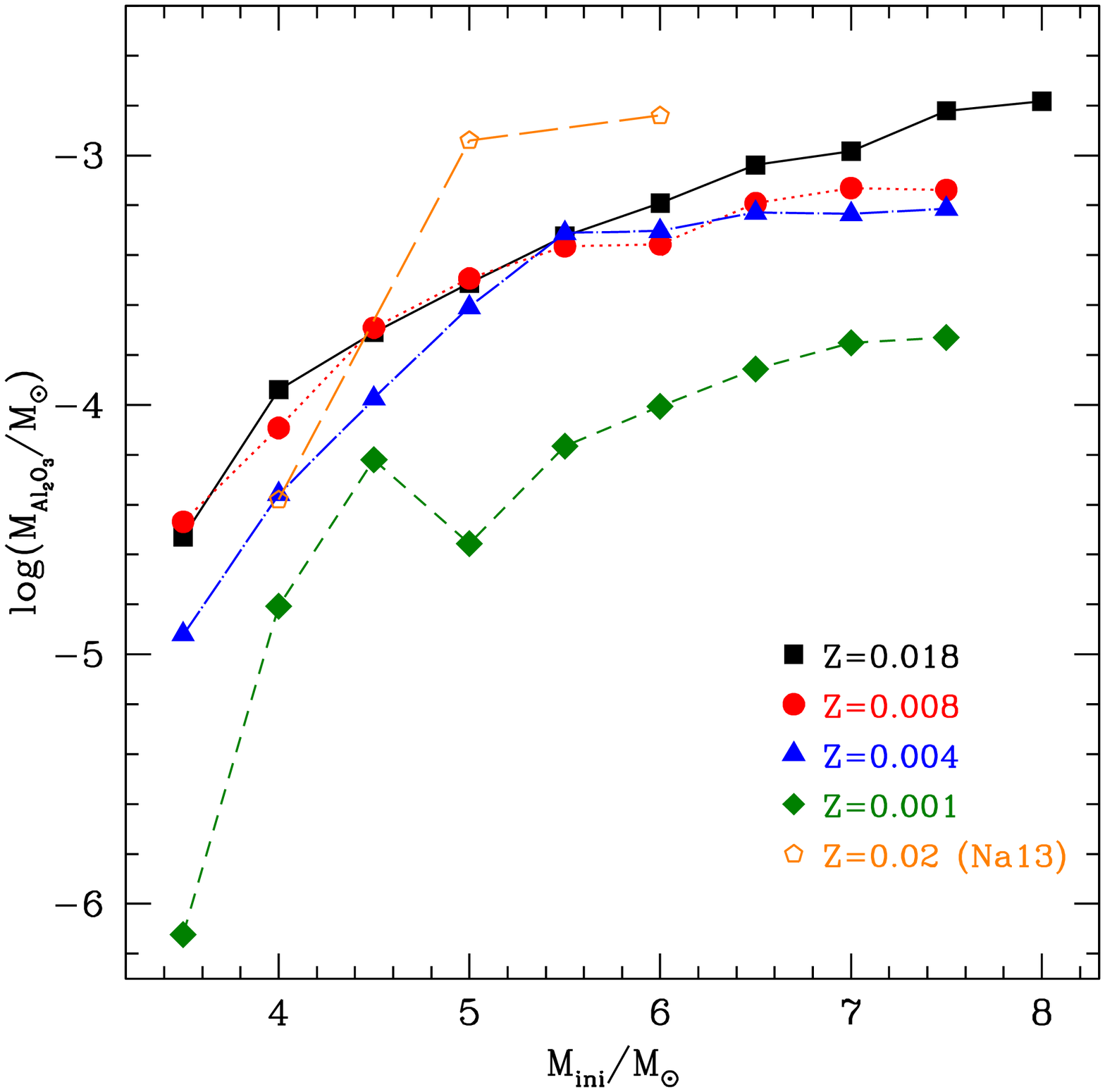}}
\end{minipage}
\vskip+10pt
\caption{The mass of silicates (left panel) and alumina dust (right panel)
produced by solar chemistry AGB stars of different mass is indicated with full
squares and connected with a solid line. Results for lower metallicity models
with $Z = 10^{-3}$, $4\times 10^{-3}$, $8\times 10^{-3}$ are indicated,
respectively, with green full diamonds (dashed line),  blue full triangles
(dotted-dashed line) and red full points (dotted line). The results from Nanni
et al. (2013, Na13) and from Ferrarotti \& Gail (2006, FG06) for $Z =0.02$ are shown,
respectively, with orange open points, connected with a long-dashed line, and
with magenta crosses, connected with a solid line (note that formation of $Al_2O_3$
particles is not described in Ferrarotti \& Gail (2006)).}
\label{fmsil}
\end{figure*}

\subsection{Dust production in massive AGB/SAGB stars}

Fig.~\ref{fsil} shows the variation of the luminosity, the size of the dust
particles formed in the wind, the fraction of the key-species (silicon and
aluminium) condensed into dust  and the dust mass-loss rate (splitted into the
silicates and the alumina dust contributions), for massive AGB/SAGB stars.

For these objects the production of dust is modulated by the luminosity 
\citep{paperI, paperII}. The larger is $L$, the larger is the mass-loss rate,
the higher is the density of the gas (see eq. 2 in \citet{ventura14}). This favours dust
production, due to the larger number of gaseous molecules available to
form dust. 

\begin{table*}
\caption{Dust mass produced by low-mass AGB and massive AGB/SAGB models of solar
metallicity. The initial stellar mass $M_{\rm ini}$ is shown in the first column. The
total mass of dust, $M_{\rm d}$ and the mass of the several dust species:
olivine ($M_{\rm ol}$), pyroxene ($M_{\rm py}$), quartz ($M_{\rm qu}$),  alumina
dust($M_{\rm Al_2O_3}$), solid iron ($M_{\rm ir}$), solid carbon ($M_{\rm C}$) 
and silicon carbide ($M_{\rm SiC}$) are also shown. All masses are
expressed in solar units.}
\label{dustmass}
\begin{tabular}{cccccccccc}
\hline
\hline
$M_{\rm ini}$ & $M_{\rm d}$ & $M_{\rm ol}$ & $M_{\rm py}$ & 
$M_{\rm qu}$ & $M_{\rm Al_2O_3}$ & $M_{\rm ir}$ & $M_{\rm C}$ & $M_{\rm SiC}$  \\
\hline
1.5  & 5.11e-04 & 8.45e-05 & 2.95e-05 & 9.40e-06 & 4.42e-08 & 1.16e-04 & 5.61e-05 & 2.08e-04  \\
1.75 & 7.17e-04 & 2.81e-05 & 1.18e-05 & 7.67e-06 & 1.72e-08 & 2.74e-04 & 2.82e-04 & 1.05e-04  \\ 
2.0  & 2.34e-03 & 6.36e-06 & 2.36e-06 & 8.71e-07 & 2.18e-09 & 1.28e-03 & 9.41e-04 & 1.10e-04  \\
2.25 & 3.23e-03 & 1.81e-06 & 7.05e-07 & 4.31e-07 & 1.31e-09 & 9.69e-05 & 2.53e-03 & 6.01e-04  \\
2.5  & 4.19e-03 & 1.62e-07 & 6.06e-08 & 2.35e-08 & 2.29e-10 & 2.29e-05 & 3.27e-03 & 8.93e-04  \\
3.0  & 6.05e-03 & 5.24e-07 & 2.08e-07 & 8.84e-08 & 5.04e-10 & 2.07e-05 & 5.09e-03 & 9.40e-04  \\
3.5  & 2.37e-03 & 1.53e-03 & 4.44e-04 & 5.35e-05 & 2.96e-05 & 2.96e-04 &    -     &    -      \\
4.0  & 3.20e-03 & 2.33e-03 & 6.00e-04 & 3.96e-05 & 1.15e-04 & 1.09e-04 &    -     &    -      \\
4.5  & 3.74e-03 & 2.75e-03 & 6.74e-04 & 3.63e-05 & 1.95e-04 & 8.11e-05 &    -     &    -      \\
5.0  & 4.48e-03 & 3.31e-03 & 7.70e-04 & 3.28e-05 & 3.07e-04 & 6.19e-05 &    -     &    -      \\
5.5  & 5.34e-03 & 3.91e-03 & 8.70e-04 & 2.95e-05 & 4.75e-04 & 5.08e-05 &    -     &    -      \\
6.0  & 6.16e-03 & 4.47e-03 & 9.61e-04 & 2.92e-05 & 6.43e-04 & 5.08e-05 &    -     &    -      \\
6.5  & 7.72e-03 & 5.60e-03 & 1.12e-03 & 2.61e-05 & 9.15e-04 & 5.74e-05 &    -     &    -      \\
7.0  & 8.20e-03 & 5.92e-03 & 1.16e-03 & 2.62e-05 & 1.04e-03 & 5.82e-05 &    -     &    -      \\
7.5  & 1.05e-02 & 7.51e-03 & 1.36e-03 & 2.20e-05 & 1.51e-03 & 7.98e-05 &    -     &    -      \\
8.0  & 1.14e-02 & 8.22e-03 & 1.43e-03 & 2.07e-05 & 1.65e-03 & 8.41e-05 &    -     &    -      \\
\hline
\end{tabular}
\end{table*}

As shown in the top-left panel of Fig.~\ref{fsil}, the luminosity 
increases during the initial AGB phases due to the increase in the core mass.
This trend is reversed at a given stage during the AGB evolution because of the
progressive consumption of the external mantle, which favours the general cooling
of the whole external structure. This behaviour reflects into the
amount and the size of the dust grains produced during the AGB life. The
silicates grains with the largest size are formed during the highest
luminosity phases (see the top-right panel in Fig.~\ref{fsil})\footnote{For
clarity sake, we show only the dominant silicate species, i.e. olivine. We discuss the properties 
of the other dust species in the text}. This is the phase where we expect  massive
AGB/SAGB stars to show the strongest IR emission, as this is sensitive to the
amount of dust in the circumstellar envelope.

Because higher mass stars evolve at larger luminosities, olivine grains
with the largest size, $a_{\rm ol} \sim 0.15 \mu$m, are produced by AGB stars with
initial mass close to the threshold to undergo core collapse, i.e. $M_{\rm ini} \sim
8~M_{\odot}$. Generally speaking, most of the olivine grains formed in the
wind of massive AGB/SAGB stars have dimensions in the range $0.1\mu$m$ < a_{\rm ol}
< 0.15 \mu$m, while pyroxene and quartz grains can be as small as
$\sim0.07\mu$m and $<0.05 \mu$m, respectively. 

The fraction of silicon condensing into solid particles (see the bottom-left panel in 
Fig.~\ref{fsil}) is $[20 - 40]\%$. In the case of stars with initial mass $M_{\rm ini}>4M_{\odot}$ these percentages are significantly smaller compared
to the alumina dust case, during almost the entire AGB phase. This is because of the large extinction coefficient of silicates, which
favours a significant acceleration of the wind as soon as silicate particles
begin to form. Under these conditions, mass continuity 
leads to a sudden drop in the density of the gas, which prevents further
formation of dust. \\

Massive AGB/SAGB stars, particularly during the highest luminosity phases, eject
great quantities of silicates into the interstellar medium (see bottom-right panel
of Fig.~\ref{fsil}). The rate at which silicate dust is ejected is in the range
$10^{-8} \,  M_{\odot}/{\rm yr} < \dot M_{\rm sil} < 10^{-6} \, M_{\odot}/{\rm yr}$.

The behaviour of alumina dust is different from silicates. Al$_2$O$_3$ is more thermodynamically stable \citep{sharp90}, thus it forms closer to the
stellar surface \citep{flavia14a}. This compound is rather transparent to the
electromagnetic radiation, so no significant acceleration of the gas occurs when
alumina dust is formed. Therefore, it is possible that high fractions of the
aluminium available condense into dust. This is shown in the
bottom-left panel of Fig.~\ref{fsil}, where it can be seen that during the
largest luminosity phases of SAGB stars ($6~M_{\odot} \leq M_{\rm ini} \leq
8~M_{\odot}$) more than $\sim 80\%$ of aluminium is condensing into
dust. The largest size of alumina dust particles, $a_{\rm Al_2O_3} \sim 0.07 \mu$m
(see top-right panel in Fig.~\ref{fsil}) are reached during the phases when all 
the aluminium available is locked into Al$_2$O$_3$ grains. The
alumina dust contribution to the total dust production rate is approximately one
order of magnitude smaller than the silicates, with Al$_2$O$_3$
mass-loss rates of $\dot M_{\rm Al_2O_3} < 10^{-7} M_{\odot}/{\rm yr}$ in all cases (see
the bottom-right panel in Fig.~\ref{fsil}).

When comparing the formation process of silicates and alumina dust,
it is important to remark that the main limiting factor to the growth of silicates
is the large wind acceleration caused by radiation pressure, that leads to 
a sudden drop of the gas density.
Conversely, the main limiting factor to the growth of alumina grains
is the amount of aluminium present in the surface regions of the star.

Alumina dust starts to condense at distances below $\sim 3\,R_{\star}$ from the stellar surface
whereas silicates particles form in a region between 4 and 9 $R_{\star}$. The
location of the condensation zone of silicates is sensitive to the initial mass
of the star, being closer to the stellar surface the lower is $M_{\rm ini}$. In
$[6 - 8]~M_{\odot}$ SAGB stars, the formation of alumina dust in an internal zone of
the circumstellar envelope, despite not sufficient to accelerate the wind,
produces a steep gradient of the optical depth (see eq. 4 in \citet{ventura14}). This
increases the gas temperature (eq. 3 in \citet{ventura14}), moving the silicates dust
condensation zone to more external regions (between $6$ and $9 \,R_{\star}$ from the stellar surface). This effect is negligible in massive AGB stars with 
$M_{\rm ini} \lesssim 5~M_{\odot}$ and the silicates dust formation takes place in a more
internal region (between 4  and 6 $R_{\star}$). \\

The masses of silicates and alumina dust produced by solar metallicity AGB stars
are listed in Table 2. In Fig.~\ref{fmsil} they are compared with results from lower-$Z$ AGB models 
\citep{paperI, paperII,dicrisci13, ventura14}. We find that
the amount of silicates and alumina dust increases with the initial mass of the
star. The mass of produced silicates dust ranges from $M_{\rm sil} \sim 2\times
10^{-3}~M_{\odot}$ ($M_{\rm ini}=3.5~M_{\odot}$) to $M_{\rm sil} \sim 9.7\times
10^{-3}~M_{\odot}$ ($M_{\rm ini}=8~M_{\odot}$). The amount of formed alumina dust 
spans the mass interval  $3\times 10^{-5} < M_{\rm Al_2O_3} < 1.7\times
10^{-3}~M_{\odot}$. Considering both dust contributions (alumina and
silicates), we find for the present models a dust-to-gas ratio ($\delta$) of
${\rm log}(\delta)\sim -3$. As discussed earlier in this section, the mass-loss rate
is the key parameter determining the overall dust production in massive AGB/SAGB
stars. 
This is the
reason why $\delta$ turns out to be only mildly dependent on $M_{\rm ini}$.

The left panel of Fig.~\ref{fmsil} shows that the mass of silicates scales almost linearly 
with $Z$. This is caused by the larger amount of silicon available in higher metallicity stars.
The trend of the alumina dust mass with metallicity is still positive but less
straightforward, the results being less sensitive to $Z$. 
The higher efficiency of HBB in lower metallicity models
favours alumina dust production (i.e., more gaseous Al is available to form
Al-based dust grains) and partially compensates the effect of the lower
metallicity. On the other hand, the exhaustion of the gaseous aluminium
available for the condensation into dust sets an upper limit to the amount of
Al$_2$O$_3$ formed. 

The present results indicate that massive AGB stars experiencing HBB are the main manufacturers of alumina dust and silicates. This is nice agreement with the recent findings by \citet{lugaro17}: comparing the $^{17}O/^{18}O$ isotopic ratio in presolar grains with the new $^{17}O(p,\alpha)^{14}Na$ proton-capture rate, they found that massive AGB stars experiencing HBB are the most likely responsible of Al-rich oxides and silicates grains formation.

\subsection{Dust production in low-mass AGB stars}

Low-mass AGB stars produce negligible quantities of silicates and alumina
dust grains during the initial part of their AGB evolution, before becoming
carbon stars. This is due to the low mass-loss rate and luminosity experienced.
The dust formed in the circumstellar envelope of these stars is mainly produced during 
the C-rich phase and is composed of solid carbon and SiC.

While luminosity is the main factor affecting dust formation in massive AGB/SAGB stars,
the formation of dust in low-mass stars is guided by the amount of carbon available
in the envelope. Therefore, the quantity of dust formed increases during the C-star phase, as
more and more carbon is transported to the stellar surface, owing to the effects
of repeated TDU episodes. The increase in the surface carbon is accompanied by an
increased cooling of the external regions, which leads to a significant decrease
in the effective temperature and to an increase in the mass-loss rate
\citep{marigo02, vm09, vm10}. This additional factor, for the reasons outlined
in the previous sections, further favours the formation and growth of C-based
dust particles.

\begin{figure*}
\begin{minipage}{0.45\textwidth}
\resizebox{1.\hsize}{!}{\includegraphics{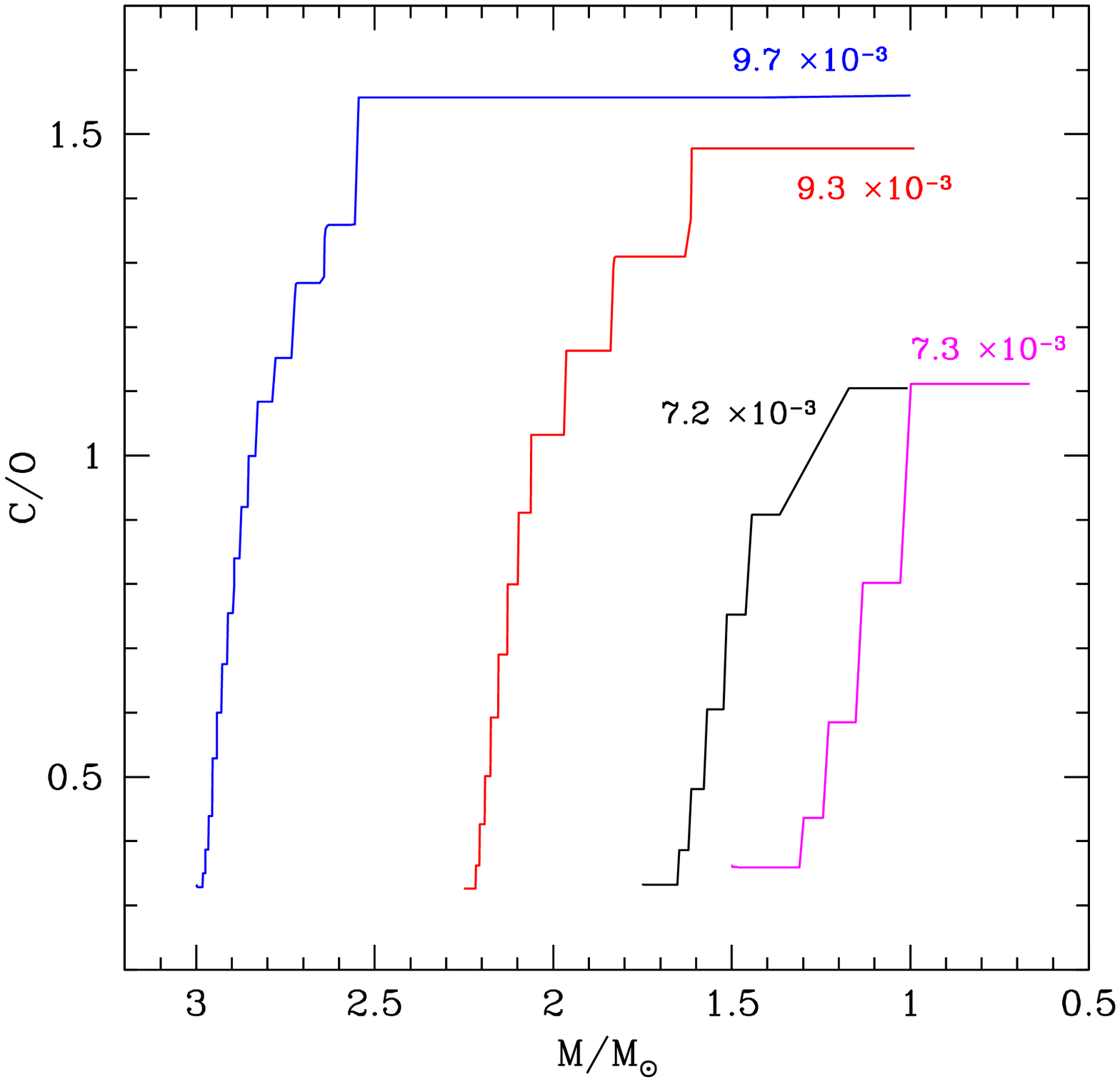}}
\end{minipage}
\begin{minipage}{0.45\textwidth}
\resizebox{1.\hsize}{!}{\includegraphics{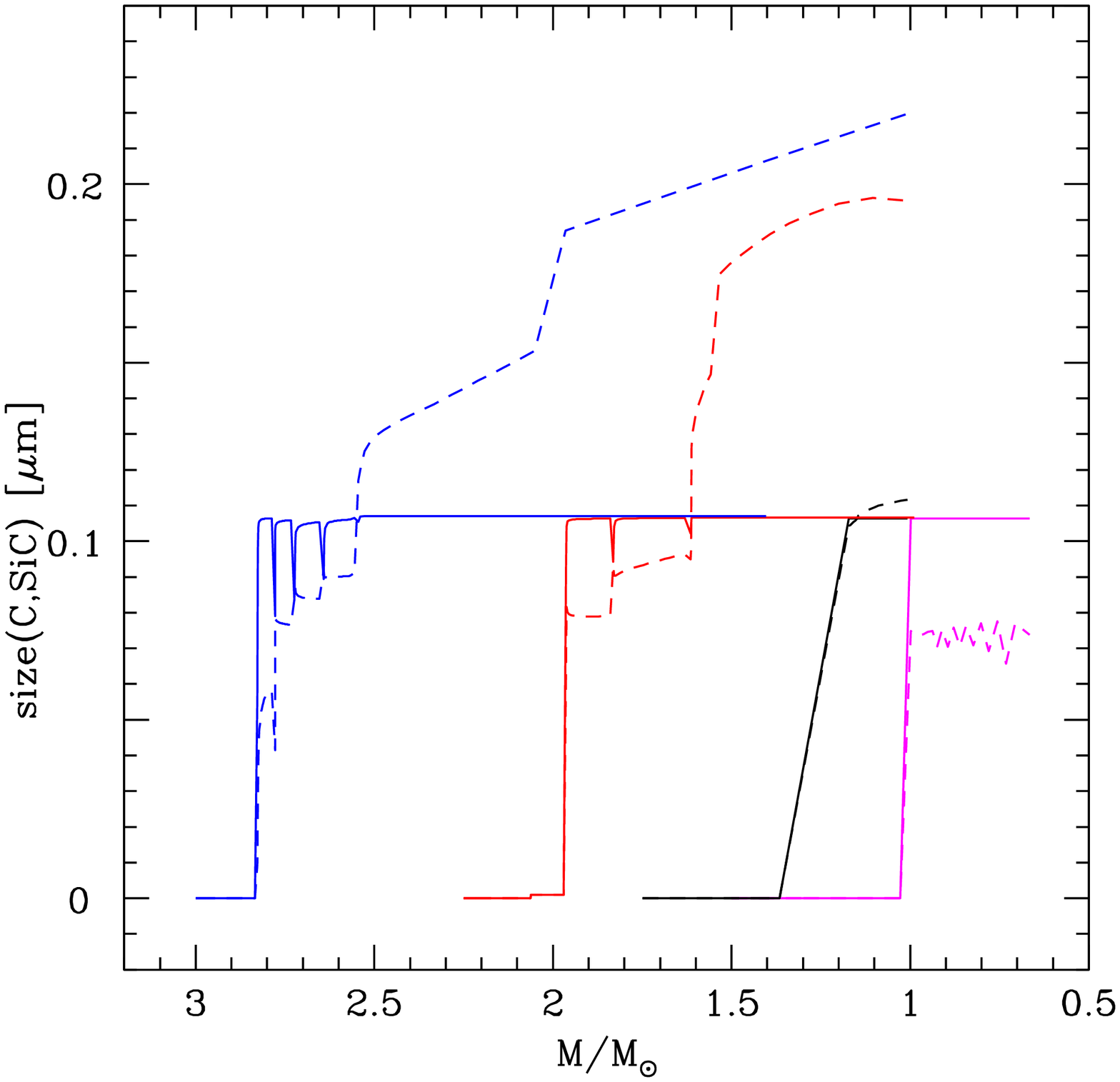}}
\end{minipage}
\vskip-0pt
\begin{minipage}{0.45\textwidth}
\resizebox{1.\hsize}{!}{\includegraphics{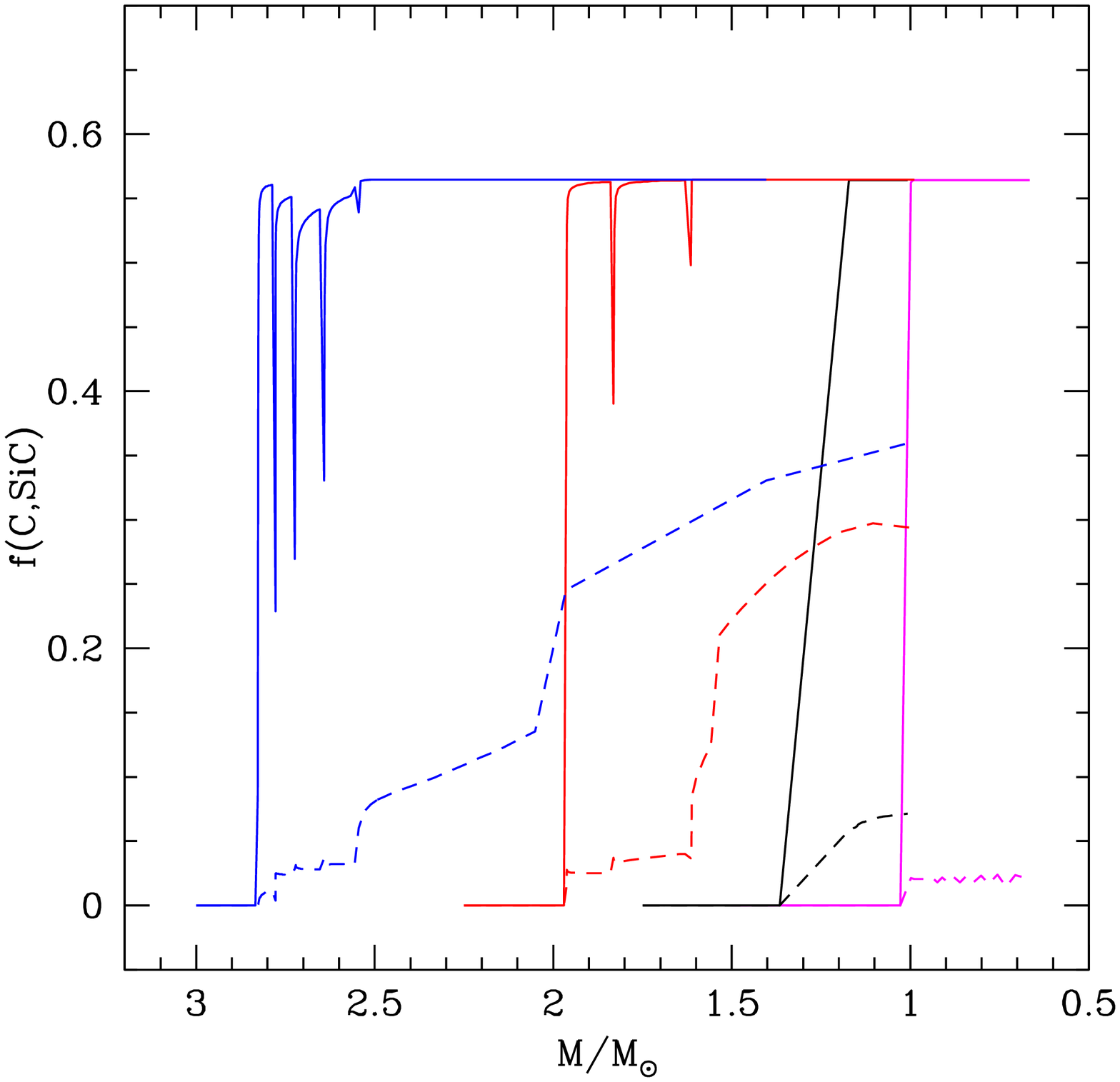}}
\end{minipage}
\begin{minipage}{0.45\textwidth}
\resizebox{1.\hsize}{!}{\includegraphics{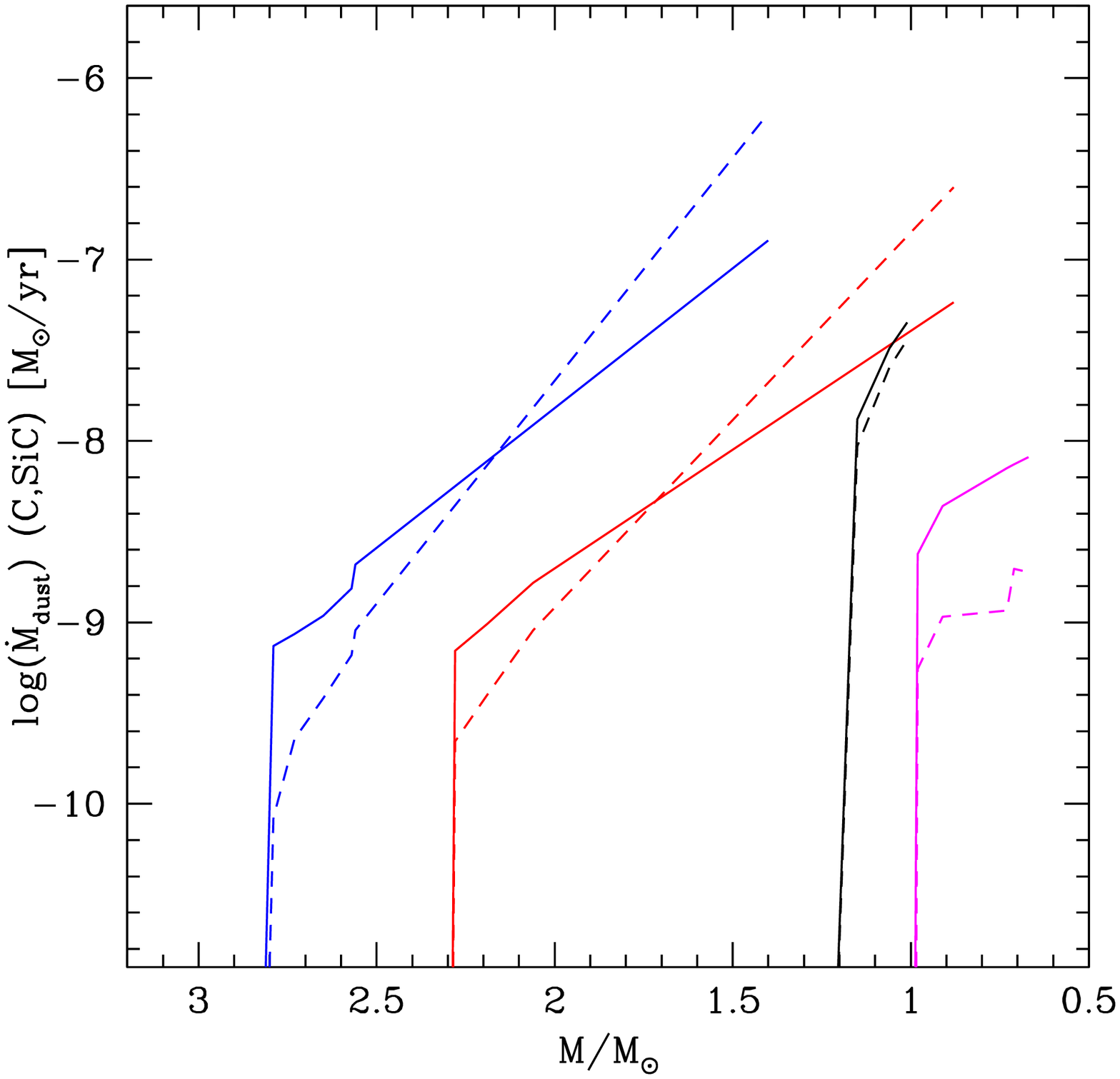}}
\end{minipage}
\vskip+10pt
\caption{Variation during the AGB phase of physical and dust properties
of stars with initial mass in the range $[1.5 - 3]~M_{\odot}$ shown by the coloured
lines from magenta to blue. The current mass of
the star (decreasing during the evolution) is displayed on the abscissa. The
quantities shown in the four panels are:  surface
C/O ratio, with the values of the final surface carbon mass fraction also
indicated (top-left), size of SiC and solid carbon grains (top-right, solid and dashed
lines, respectively),  fraction of silicon condensed into SiC
and of carbon condensed into dust (bottom-left, solid
and dashed lines, respectively), SiC and carbon mass-loss rates (bottom-right, solid and dashed lines, respectively).
A coloured version of this plot is available online.}
\label{fcarb}
\end{figure*}

Fig.~\ref{fcarb} shows the variation of the surface C/O ratio, the size of 
carbon and SiC dust grains formed, the fraction of carbon and silicon condensed
into dust, and the mass loss rate of carbon and SiC dust,
during the AGB phase of stars with initial mass $1.5~M_{\odot} \leq M_{\rm ini}
\leq 3~M_{\odot}$.

The size of the carbon dust grains increases during the AGB phase (see top-right
panel) as a consequence of the increase in the surface
carbon abundance (top-left panel). Therefore,  carbon grains with the
largest size form during the final AGB phases, when the surface carbon mass
fraction reaches the maximum value. The typical size of carbon grains
is in the range $0.05\mu$m$ < a_{\rm C} < 0.25 \mu$m, and it increases
with the initial stellar mass, because stars of
higher mass experience a higher number of TDU episodes, accumulating a higher
abundance of carbon in the external layers (DC16). This is clear from
Fig.~\ref{fcarb} (top-right panel) where models with $M_{\rm ini}<2 \, M_{\odot}$, whose
surface C/O barely exceeds unity, form smaller dust grains than their more
massive counterparts.

The fraction of gaseous carbon condensed into dust, shown in the botton-left
panel of  Fig.~\ref{fcarb}, ranges from $\sim 10\%$ to $\sim 35\%$. The upper
limit is motivated by the large values of the carbon dust extinction coefficients, 
which favour a rapid acceleration of the wind once
carbon grains begin to form. This is similar to the effect on the dynamics of
the wind triggered by the formation of silicates in the circumstellar envelope
of massive AGB/SAGB stars.

A different behaviour is found for SiC. Although SiC is more stable
and it is formed in more internal 
circumstellar regions (with higher density), the growth of SiC particles 
never exceeds the threshold
value of $a_{\rm SiC} \sim 0.1\mu$m (see the top-right panel in Fig.~\ref{fcarb}),
which corresponds to the situation where $\sim 55\%$ of gaseous silicon is
condensed into dust (see bottom-left panel in Fig.~\ref{fcarb}). This is the
largest amount of silicon available to form dust, because of the high stability
of the SiS molecule, which makes $\sim 45\%$ of the total  silicon in the
envelope to be locked into gaseous SiS \citep{fg06}.  

The region where SiC forms is $\sim [3.5 - 5.5] R_{\star}$ from the surface of the star, to be compared 
with the larger distances, $\sim [5.5 - 9] R_{\star}$, of the carbon condensation zone. \\
Similarly to what
found for oxygen-rich stars, the formation of the most stable
dust species, i.e. SiC, favours the increase in the optical depth, so that solid carbon condensation 
 takes place in a more external region of the wind. This effects gets more
and more important as the stars evolves through the AGB, because the density of the
wind increases towards the final evolutionary phases.
In stars with initial mass $M_{\rm ini} <2M_{\odot}$, which achieve only a modest C/O 
during the AGB evolution, the residual gaseous carbon in the wind
after SiC formation in the internal regions of the circumstellar envelope
is such that the size of the carbon grains formed is barely equal or even smaller than
the SiC grains.

\begin{figure*}
\begin{minipage}{0.48\textwidth}
\resizebox{1.\hsize}{!}{\includegraphics{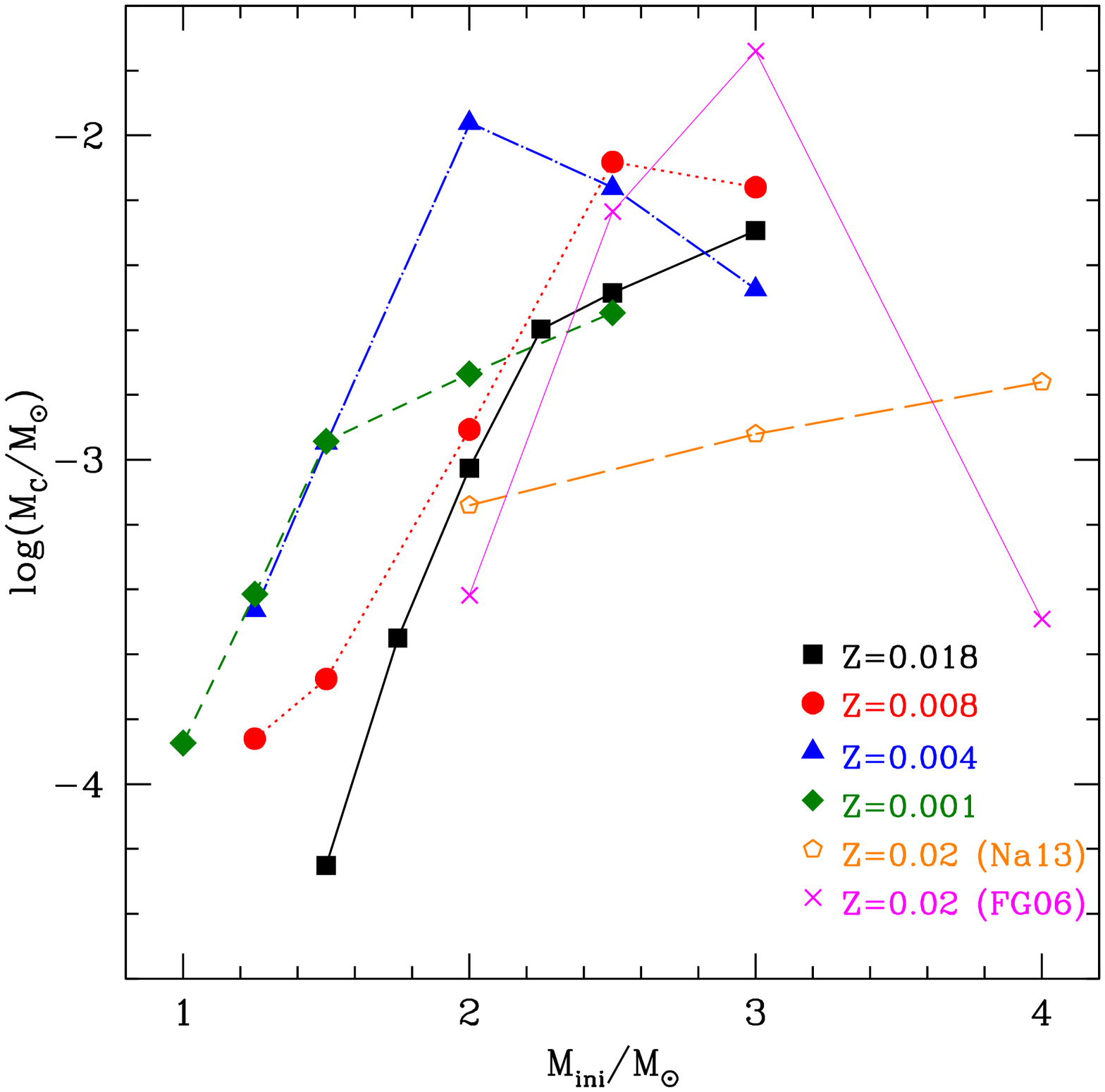}}
\end{minipage}
\begin{minipage}{0.48\textwidth}
\resizebox{1.\hsize}{!}{\includegraphics{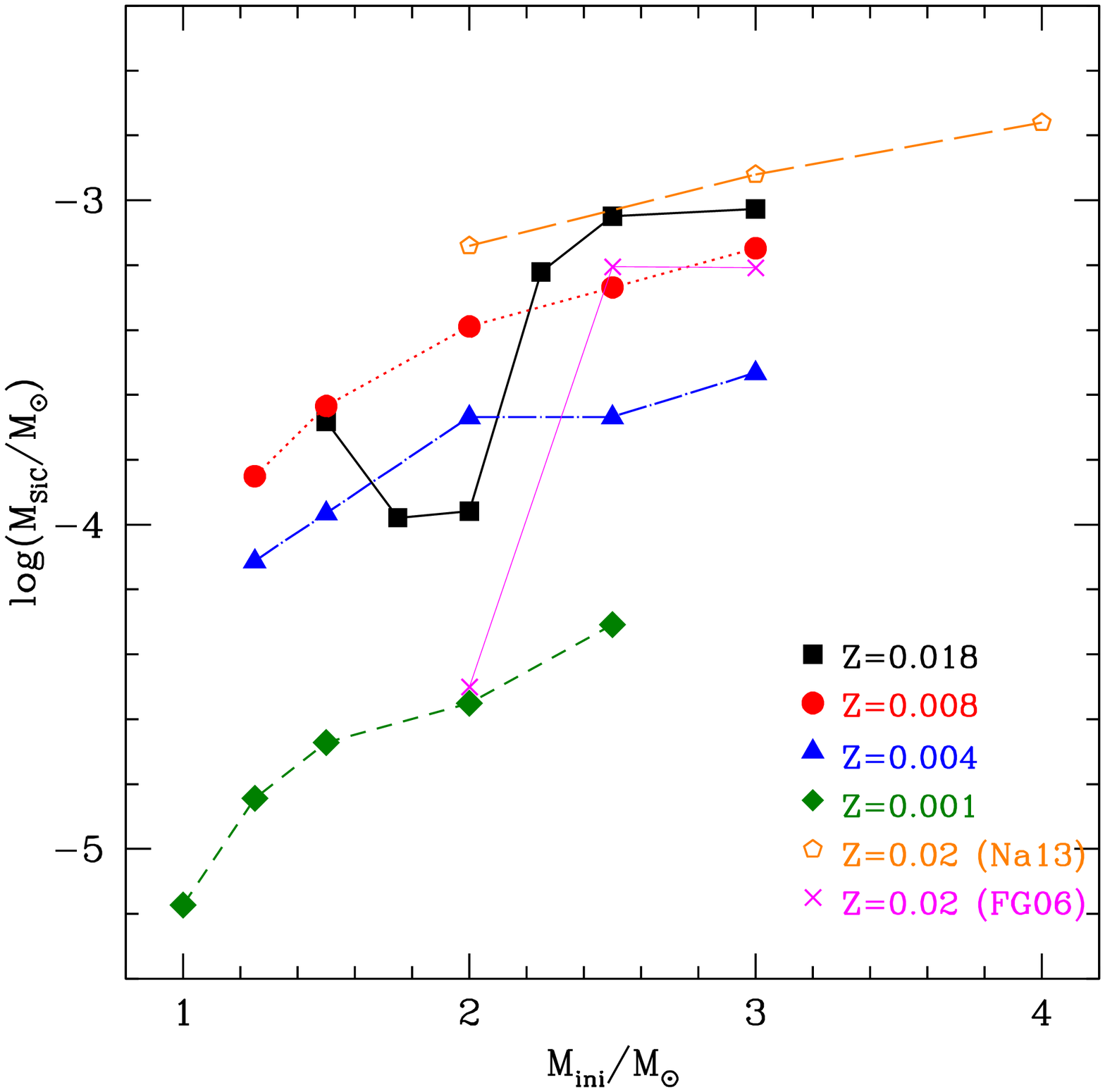}}
\end{minipage}
\vskip+10pt
\caption{The mass of solid carbon (left panel) and silicon carbide (right panel)
produced by solar chemistry AGB stars of different mass compared with models of similar
metallicity by Nanni et al. (2013, Na13) and Ferrarotti \& Gail (2006, FG06) and with 
the results  for lower metallicity models ($Z = 10^{-3}$, 4$\times 10^{-3}$ and 
8$\times 10^{-3}$). The same symbols as in Fig.~\ref{fmsil} are adopted.}
\label{fmcarb}
\end{figure*}

The rates at which carbon and SiC dust are ejected into the interstellar medium increase during the AGB evolution. 
The largest values, attained in the final evolutionary phases, are of the order of $\dot M_{\rm C} \sim
10^{-6} \dot M/{\rm yr}$ for carbon dust and $10^{-8} \dot M /{\rm yr} < \dot M_{\rm SiC} <
10^{-7} \dot M/{\rm yr}$ for SiC (see the bottom-right in Fig.~\ref{fcarb}).

The masses of carbon and SiC dust produced by solar metallicity low-mass AGB
stars are shown in Fig.~\ref{fmcarb} and compared with the corresponding dust
masses from lower-metallicity stars.
The mass of carbon dust is in the range $10^{-4}~M_{\odot} < M_{\rm C} <
10^{-2}~M_{\odot}$. The trend with $M_{\rm ini}$ is positive, as
expected (see the reasons discussed above). The trend with metallicity, however,
is not trivial. For stars with $M_{\rm ini} \leq 2~M_{\odot}$, the amount of
carbon dust produced by solar metallicity models is smaller than what
found at lower $Z$. Because of the higher initial oxygen, the C-star stage is attained
in a more advanced AGB phase and the carbon excess with respect to oxygen is
generally smaller. For stars with $M_{\rm ini} \sim [2.5 - 3]~M_{\odot}$, this is
compensated by the larger temperatures at the base of the envelope attained by
lower metallicity stars of the same mass, which favours the ignition of a soft
HBB, preventing the accumulation of high quantities of carbon in the surface
regions.

The dust-to-gas ratio of carbon-stars ranges from
${\rm log}(\delta)~-3$, at the beginning of the carbon-rich phase, up  
${\rm log}(\delta)~-2.4$, achieved during the final phases of AGB stars with 
$M_{\rm ini}=[2.5 - 3]~M_{\odot}$.

The mass of SiC produced is in the range 
$10^{-4}~M_{\odot} < M_{\rm SiC} < 10^{-3}~M_{\odot}$. Although the abundance of the key species (silicon in this case) results to be proportional to the metallicity, 
this is not the case for $M_{\rm SiC}$.  For $M_{\rm ini} \leq 2~M_{\odot}$, $M_{\rm SiC}$ is
comparable to what is found for the $Z = (4 - 8) \times 10^{-3}$ models, whereas for
higher masses the SiC dust mass for $Z = 0.018$ is slightly larger than for lower-$Z$ 
AGB stars. The reasons for this behaviour are the same as
those illustrated above to explain the trend of carbon dust with metallicity. In addition, a
significant fraction of the mass of the envelope is lost when the stars are
still oxygen-rich, thus the mass lost by the star during the phase when SiC dust
is formed is smaller. This effect partly counterbalances the higher silicon
abundance present in the star.

For what concerns the relative contributions of solid carbon and SiC to the
overall dust production by low-mass AGB stars, we find that the individual
contributions of the two species are sensitive to the initial mass of the star.
Stars with $M_{\rm ini} \lesssim 2~M_{\odot}$ become carbon stars only in the
very final AGB phases, with a surface C/O only slightly above unity (DC16; see
also the top-left panel of Fig.~\ref{fcarb}). In these conditions, the excess of
carbon with respect to oxygen is so small, that the masses of SiC and solid
carbon are very similar. On the other hand, in stars with $M_{\rm ini} >
2~M_{\odot}$, we find that C/O$ \sim 1.5$ and the amount of carbon not
locked into CO molecules is sufficiently large to allow the formation of solid
carbon in excess of SiC.

\subsection{Comparison with other AGB dust yields}
In this section we focus our attention on the comparison with different results on the 
AGB dust yields (i.e., the mass of the various dust species produced), obtained using the 
same methodology for the dust formation model but a different AGB phase description. 

As stated previously, we use the same approach introduced for the first time by 
Ferrarotti \& Gail (2006, hereinafter FG06) to describe the wind dynamics and dust production. 
FG06 base their modelling on a synthetic description of the AGB phase. The metallicity used 
by FG06 ($Z=0.020$) is slightly higher than the value adopted here ($Z=0.018$) but the 
difference is sufficiently small to allow a straight comparison of the results.

In Figs.~\ref{fmsil} and \ref{fmcarb} we show FG06's dust yields, to allow a direct 
comparison with our results. The different parametrisation adopted to describe the TDU 
efficiency is likely the reason for the difference in the minimum mass reaching the
C-star stage, which is 2$M_{\odot}$ in FG06, compared to 1.5$M_{\odot}$, found in the
present work, for models of solar metallicity. A common behaviour of both sets of models 
is the increase in the amount of solid carbon and SiC produced with the initial mass.
In the FG06 case larger quantities of solid carbon are found in the range of mass
close to the lower threshold required to activate HBB; this is probably due to the larger 
number of thermal pulses experienced in the FG06 models, resulting in a larger surface 
carbon available to form carbon dust. Stars with initial masses lower than $2M_{\odot}$ 
do not become carbon stars, and are responsible for production of silicates dust in the range 
$10^{-3}-10^{-4} M_{\odot}$. 

For what concerns the massive star domain, FG06 assume that the HBB is active for
$M \geq 4M_{\odot}$. While the production of silicates is comparable with our results in 
the lower limit of this range of masses, for the most massive AGB stars the amount of
silicates formed is a factor $\sim 2$ smaller that in the present models: this is due to
the higher efficiency of the HBB process, in turn determined by the larger efficiency of the
convective models adopted here. An additional difference between the present results and
FG06 in the dust budget of massive stars is that in the latter models some carbonaceous 
dust is produced at the end of the AGB phase, owing to the effects of some late TDU events,
after HBB is extinguished.

The same methodology introduce by FG06, and used in the present work, was used also by Nanni et al. (2013, hereafter Na13), 
to calculate the dust yields by AGB stars with solar
chemistry. For the sake of clarity, we compare silicates total dust production to the 
low condensation temperature case (i.e. chemical sputtering case) adopted by Na13, which is 
the same approximation considered in the present work. 
Yet, Na13 state that low and high condensation temperature approaches 
reach approximately the same condensation degrees during the phase that dominate the total 
dust production.

The dust yields predicted by Na13 are shown in
Figs.~\ref{fmsil} and \ref{fmcarb}. 
An obvious difference is the threshold mass
that separates stars producing mainly carbonaceous dust from those producing
silicates and alumina dust. This mass limit is $M_{\rm ini} \sim 3~M_{\odot}$ in the present 
computation, while it is $M_{\rm ini} \sim 4~M_{\odot}$ in Na13. \\

This
dissimilarity stems from the different convection modelling. Our AGB
computations are based on the Full Spectrum of Turbulence (FST) convective model
\citep{cm91}. As shown by \citet{vd05}, the use of the FST modelling leads to
very efficient HBB conditions, i.e. the minimum mass required to experience HBB
in the FST model is significantly lower than for other AGB models in the literature
(e.g., Karakas \& Lugaro 2016).

For massive AGB stars with $M_{\rm ini}$ in the range $[5 - 6]~M_{\odot}$, the amounts
of alumina and silicates dust predicted by Na13 are approximately a factor of 2
higher than in our AGB models (see Fig.~\ref{fmsil}). This difference is
mainly due to the different duration of the largest luminosity phase that is
coupled with the maximum dust production phase. In Na13, this phase is a factor
of $\sim 2$ longer than in our AGB models. In the case of alumina dust, the
choice of the sticking coefficients also affect this difference. We use 
$\alpha_{\rm Al_2O_3}=0.1$, while Na13 adopted a much higher value of
$\alpha_{\rm Al_2O_3}=1$. 

For low-mass AGB stars, the mass of SiC is similar in the two studies, while the
amount of carbon dust produced by our models is approximately a factor 2 larger
than Na13. This is likely due to the different choice adopted for the fraction of seed
nuclei with respect to the number density of hydrogen molecules ($\epsilon_s$).
We follow \citep{fg06} and assume that $\epsilon_s=10^{-13}$, while Na13 assume
that the number of seed particles relevant for the formation of carbon grains
scales with C/O.

\subsection{The total dust production from solar metallicity AGB stars }

\begin{figure}
\begin{minipage}{0.50\textwidth}
\resizebox{1.\hsize}{!}{\includegraphics{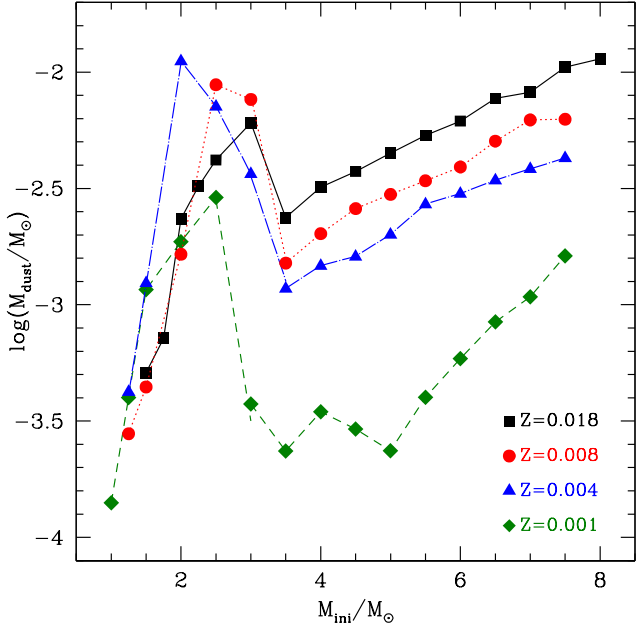}}
\end{minipage}
\caption{The total dust mass produced by solar metallicity AGB models of
different initial mass are indicated with black, full squares and are connected
with a solid line. Also shown are results of lower-metallicity models, with
the same symbols used in Fig.~\ref{fmsil}. }
\label{fdust}
\end{figure}

Fig.~\ref{fdust} shows the total dust mass produced by AGB stars of solar
metallicity. In order to better understand the trend of the dust mass with
metallicity, we also show the results from our previous works focused on
lower-$Z$ AGB stars \citep{paperI, paperII, dicrisci13, ventura14}.

The dependence of the dust mass produced on the initial stellar
mass can be understood on the basis of the arguments
presented earlier in this section and can be summarised as follows:

\begin{enumerate}

\item{In low-mass AGB stars, the total amount of dust produced ($M_{\rm dust}$),
composed of solid carbon and SiC, increases with the initial stellar mass. We
find $M_{\rm dust} \sim 5 \times 10^{-4}~M_{\odot}$ for  $M_{\rm ini} = 1.5~M_{\odot}$,
up to $M_{\rm dust} \sim 6\times 10^{-3}~M_{\odot}$ for  $M_{\rm ini} = 3~M_{\odot}$.
The reason for this is that the number of TDU events experienced increases with
$M_{\rm ini}$. Consequently, higher mass stars accumulate more carbon in
the external regions, which favours the formation of carbon dust.}

\item{For what concerns the dust composition, we find that for increasing
$M_{\rm ini}$ the ratio between the mass of solid carbon and the mass of SiC becomes
larger. This is because the amount of SiC formed, unlike carbon, is rather
independent of the surface chemical composition, as the most relevant species
for the formation of SiC is silicon.}

\item{The AGB model forming the minimum amount of silicate dust, $M_{\rm dust} \sim
2.4\times 10^{-3}~M_{\odot}$, has an initial mass $M_{\rm ini} = 3.5~M_{\odot}$. 
This is the lowest mass experiencing HBB, which destroys surface carbon,
thus preventing the formation of carbonaceous dust. On the other hand, the
production of silicates is small, because this star experiences a soft HBB
during the AGB phase and the luminosity is below $25,000 L_{\odot}$. }

\item{For massive AGB/SAGB stars, the mass of dust formed increases 
with $M_{\rm ini}$, ranging from the
quantity given in the previous point to $M_{\rm dust} = 1.14\times
10^{-2}~M_{\odot}$, for the most massive SAGB star with $M_{\rm ini} = 8~M_{\odot}$.
This is because higher mass stars experience a
stronger HBB, thus evolving at larger luminosities and mass-loss rates. }

\item{For massive AGB/SAGB stars the majority of the dust formed is composed of
silicates, with $\sim 10\%$ of alumina dust.} 
\end{enumerate}

The dependence of dust production on the initial metallicity of massive AGB/SAGB stars is rather
straightforward. Dust production increases with $Z$, because the
initial mass fractions of the key elements relevant to the formation of
silicates and alumina dust (i.e. silicon and aluminium, respectively) increase
almost linearly with $Z$, and their abundance does not change during the AGB
evolution. The trend of the dust mass with $Z$ is not completely linear though,
because lower-$Z$ AGB stars experience a more efficient HBB, which, for the
reasons given above, favours a more efficient formation of silicates and of
$\rm Al_2O_3$.

For low-mass AGB stars, the dependence of dust production on the initial stellar metallicity
 is more tricky. In these
stars most of the dust formed is composed of solid carbon, which is sensitive to
the amount of carbon accumulated in the surface regions, as a consequence of
repeated TDU episodes. Solar metallicity AGB stars produce, on average, a
smaller quantity of dust compared to their lower metallicity counterparts.
The reasons for this are twofold: a) the C-star stage is reached later in the
AGB evolution, when a significant fraction of the envelope has been lost; b) the
initial oxygen in the star is larger, thus a smaller quantity of carbon (in
excess of oxygen) is available to form dust.

\section{Conclusions}

In this paper, we have completed the grid of ATON dust yields for AGB/SAGB stars
extending our previous calculations to solar metallicity stars. Our main results can
be summarised as follows. \\
The kind of dust
produced reflects a dichotomy (C-rich vs. O-rich) in the evolution of the
surface chemical composition, which depends on the initial stellar mass, $M_{\rm ini}$. 

In low-mass AGB stars, with $1.5~M_{\odot} \leq M_{\rm ini} \leq 3~M_{\odot}$, the surface
chemical composition is altered by the several TDU episodes that eventually lead
to the formation of a carbon star (C/O$>$1). The formation of solid carbon and
SiC grains takes place in the circumstellar envelope of these stars. The size of
SiC grains formed keeps around $a_{\rm SiC} \sim 0.1\mu$m for the whole AGB
life. However, the size of carbon grains increases with the increasing
amount of carbon being accumulated in the surface regions. The carbon dust
particles with the largest sizes, $a_{\rm C} \gtrsim 0.2\mu$m, are formed in
the very final evolutionary phases of AGB stars with $M_{\rm ini} =
2.5-3~M_{\odot}$. The mass of carbonaceous dust formed increases with the mass
of the star, ranging from a few $10^{-4}~M_{\odot}$ for $M_{\rm ini} \sim
1.5~M_{\odot}$, to $\sim 10^{-3}~M_{\odot}$ for $M_{\rm ini} \sim 2.5-3~M_{\odot}$.
In the lowest mass stars the dominant dust component is SiC, while in their more
massive counterparts most of the dust produced is under the form of solid
carbon.

Massive AGB/SAGB stars with $M_{\rm ini} > 3~M_{\odot}$ experience HBB. In
this case the formation of carbon stars is inhibited by the destruction of 
surface carbon via proton-capture reactions at the base of the convective
envelope. The most relevant dust species formed in these stars are silicates and
alumina dust. 
The latter is more stable than silicates, but the amount
of dust that can be produced is severely limited by the scarcity of aluminium.
Alumina grains grow until reaching dimensions of the order of $a_{\rm Al_2O_3} \sim
0.06-0.07\mu$m. Most of the dust produced by massive AGB/SAGB stars is under the
form of silicates. The size reached by the dominant silicates grains, i.e.
olivine, are in the range $0.1\mu m < a_{\rm ol} < 0.15\mu$m. The mass of dust
produced increases with $M_{\rm ini}$ because higher mass stars
evolve on more massive degenerate cores and are exposed to stronger HBB
conditions. We find $M_{\rm dust} \sim 3.4\times 10^{-3}~M_{\odot}$ for $M_{\rm ini}
\sim 3.5~M_{\odot}$, up to a maximum of $M_{\rm dust} \sim 1.1\times
10^{-2}~M_{\odot}$ for $M_{\rm ini} \sim 8~M_{\odot}$. Silicates are the largely
dominant dust species in all cases.

Previously computed ATON AGB models with dust formation have been able to
explain and interpret photometric observations of evolved stars in several low-metallicity 
galaxies of the Local Group.
The results of the present study will allow to extend these studies to 
evolved stellar populations in our Galaxy and another high-metallicity LG galaxies. 
In addition, the complete grid of dust and metal yields based ATON AGB/SAGB models with
metallicity $0.001 \leq Z \leq 0.018$ and initial mass $1.5 M_\odot \leq M_{\rm ini} \leq 8 M_{\odot}$
will be a valuable tool for chemical evolution studies, allowing to estimate the contribution of
intermediate mass stars to metal and dust enrichment on timescales $> 40$~Myr.
In a more general context, the solar metallicity AGB dust yields will allow to compare the relative role of AGB and supernovae (SN) as
stellar sources of dust over a wide range of metallicity, and to assess 
the contribution of AGB stars to the existing dust mass of the Milky Way, 
complementing the work that has already been done on the Magellanic Clouds.

\section*{Acknowledgments}
DAGH was funded by the Ram\'on y Cajal fellowship number RYC$-$2013$-$14182.
DAGH and FDA acknowledge support provided by the Spanish Ministry of
Economy and Competitiveness (MINECO) under grant AYA$-$2014$-$58082-P.
RS and RV acknowledge funding from the European Research Council under the European
Unions Seventh Framework Programme (FP/2007-2013)/ERC
Grant Agreement no. 306476. MDC acknowledges the contribution of the FP7 SPACE project
ASTRODEEP (Ref. No. 312725), supported by the European Commission

\end{document}